\documentclass[journal=jpcl,manuscript=article]{achemso}
\setkeys{acs}{articletitle = true}

\usepackage{amsmath}
\usepackage{graphicx}
\usepackage{amsfonts}
\usepackage{color}
\usepackage{soul}
\usepackage{wrapfig}
\usepackage{titlecaps}
\makeatletter
\def\captionof#1#2{{\def\@captype{#1}#2}}
\makeatother
\def\beq{\begin{equation}}
\def\eeq{\end{equation}}
\def\beqa{\begin{eqnarray}}
\def\eeqa{\end{eqnarray}}
\def\e{\varepsilon}

\def\cH{{\mathcal H}}

\def\cZ{{\mathcal Z}}
\def\cT{{\mathcal T}}

\newcommand{\appropto}{\mathrel{\vcenter{
			\offinterlineskip\halign{\hfil$##$\cr
				\propto\cr\noalign{\kern2pt}\sim\cr\noalign{\kern-2pt}}}}}
\newcommand{\ket}[1]{\left| #1\right\rangle}

\author{Omer Goldberg}
\affiliation{Department of Physics and the Ilse Katz Center for nanoscale Science and Technology, Ben-Gurion University of the Negev, Beer Sheva, Israel}
\author{Yigal Meir}
\affiliation{Department of Physics and the Ilse Katz Center for nanoscale Science and Technology, Ben-Gurion University of the Negev, Beer Sheva, Israel}
\author{Yonatan Dubi}
\affiliation{Department of Chemistry and the Ilse Katz Center for nanoscale Science and Technology, Ben-Gurion University of the Negev, Beer Sheva, Israel}
\email{jdubi@bgu.ac.il}
\title{Vibration-Assisted and Vibration-Hampered Excitonic Quantum Transport} %  

\begin{document}
\begin{abstract}
The interplay between exctions and vibrations is considered to be a key factor in determining the exciton transfer properties in light-harvesting exciton transfer complexes. Here we study this interplay theoretically in a model for exciton transport, composed of two chromophores coupled to an exciton source and sink in the presence of vibrations. We consider two cases, that show qualitatively distinct transport features. In the first, the vibrations are global and affect the two chromophores simultaneously. In the second case, the vibrations are localized on each chromophore. For global vibrations, the current exhibits {\sl anti-resonances} as a function of the chromophore energy difference, which are due to exciton-polaron interference. For local vibrations, on the other hand, the currents shows tunneling {\sl resonances} at multiples of the vibration energy. Counter-intuitively, both effects increase with increasing temperature. Our results demonstrate that an environment can either assist or hamper exciton transport, and are in accord with current understanding of energy transfer in natural exciton transfer complexes. 
\end{abstract}

\maketitle
%In molecular junctions, for example, electron-phonon interactions has been extensively investigated [CITE Galperin review], and numerous effects have been associated with them. Examples include (but not limited to) the appearance of inelastic transport bands (which allows for inelastic electron tunneling spectroscopy)[CITE], heating effect [CITE], electron dephasing [Fraunheim Pezzoli], quantum unfurling [Peskin], negative differential conductance [Galperin, Gauger] and more. 

\newpage
%\sec{Introduction} 
Excitons - bound electron-hole pairs - are the energy carriers in the photosynthetic process, carrying the harvested solar energy from the antenna (where it is collected) to the reaction center (where it is converted to chemical energy), through a network of proteins, so-called exciton transfer complex (ETC). Interest in the dynamics of excitons in the ETC exploded following recent experiments, where ultrafast nonlinear spectroscopy signals showed long-lived oscillations \cite{engel2007,Calhoun2009,Collini2010,Panitchayangkoon2010}.  These were attributed to an interplay between (quantum coherent) excitons and the vibrations of the proteins, which serve as an external environment. The idea that in natural photosynthetic complexes, which are extremely efficient \cite{fleming2004,levi2015}, quantum coherence in the presence of an environment is used to assist energy transfer, has generated much excitement ~\cite{Ishizaki2012,Collini2013,Lambert2013,Pachon2012} and debate \cite{kassal2013,Miller2012,ritschel2011,duan2017nature,Wilkins2015}. Since, many theoretical works demonstrated this effect, so-called "environment-assisted quantum transport" (ENAQT), in exciton transfer complexes \cite{mohseni2008,plenio2008,Rebentrost2009,chin2010,Wu2010,Wu2012,Moix2011} as well as other nanoscale quantum systems \cite{Semiao2010,Nalbach2010, Scholak2011a,Ajisaka2015,Lim2014,leon2015noise,Scholak2011a,Biggerstaff2016,viciani2016disorder,Caruso2016,sowa2017,Sowa2017a,Zerah-Harush2018}.

The majority of theoretical works, aiming at reconstructing the experimental results, start either with a very complicated environment model, or a complicated excitonic model, or both. Identifying the underlying mechanism of ENAQT is thus a challenging task, and many different suggestions for its origin have been made. Here, we examine the interplay between exciton transport and vibrations in a simple model system, in which the signatures of the environment on the transport properties can be traced to their origin. Specifically, we examine exciton transport through a bi-chromophore system coupled to an exciton source and sink, where the chromophores are coupled to monochromatic vibrations. We compare two cases; in the first, the two chromophores are coupled to a single vibrational mode (so-called {\sl global} phonon). In the second, each chromophore is coupled to its own vibrational mode (with the same frequency), and the vibrations on the two chromophores are completely uncorrelated (so-called {\sl local} phonons).

 Whether the vibrations are global or local is determined by the strength of spatial correlations between the vibrations surrounding each chromophore. These correlations are determined by the inter-chromophore distance, the protein structure etc., The distinction between local and global phononic environments and its effect on, e.g. donor-acceptor electron transfer and exction transfer in light-harvesting complexes, has been studied by several authors \cite{Struempfer2011,Nesterov2015,Nazir2009,Nalbach2010,Merkli2016,Park2011, Wu2010}, finding that the nature of the vibrations can have a decisive effect on the electron/exciton transfer rate and on coherence life-times. For instance, in Ref.~\citenum{Struempfer2011} population transfer times were shown to be shortened and absorption spectra were shown to be narrowed for the case of negatively-correlated vibrations (and the reverse for positively-correlated vibrations). In Ref.~\citenum{Wu2010} the authors showed that, depending on the reorganization energy, exciton transfer rate can either increase or decrease in the presence of spatial vibrational correlations. In Ref.~\citenum{Nalbach2010} the authors showed the decoherence rate decreases in the presence of spatial vibrational correlations. In Ref.~\citenum{Park2011} the authors studied the effect of correlations on charge transfer in molecular junctions, showing that these can either assist or impede charge transport, depending on the parameters of the junction.   

In all the studies above, the excitonic system was coupled to a continuum bath of vibrations (in similarity to the spin-boson model). Therefore, analytic results were hard to obtain and the majority of these studies were based on numerical calculations, which made it difficult to identify exactly the interplay between the vibrations and the exciton dynamics. Here we introduce models which are simple enough to allow for analytic results to be obtained. Furthermore, here we evaluate the exciton current at steady state (rather than evaluate the exciton population dynamics, which is the typical quantity calculated in the studies above). This allows us to determine directly the effect of the vibrational environment on exciton current. 

Despite the relative simplicity of the models, we find that both of them contain reach physics, manifested through unique transport signatures of the exciton-phonon coupling. Specifically, the global-phonon case exhibits {\sl anti-resonances} in the energy current due to interference effects. Qualitatively different,the global case exhibits tunneling-resonant current. Surprisingly, for both cases the non-trivial features tend to {\sl increase} with increasing temperature, in opposite to what is expected from quantum coherent effects.

\begin{figure}[h!]
\includegraphics[width=3.5in]{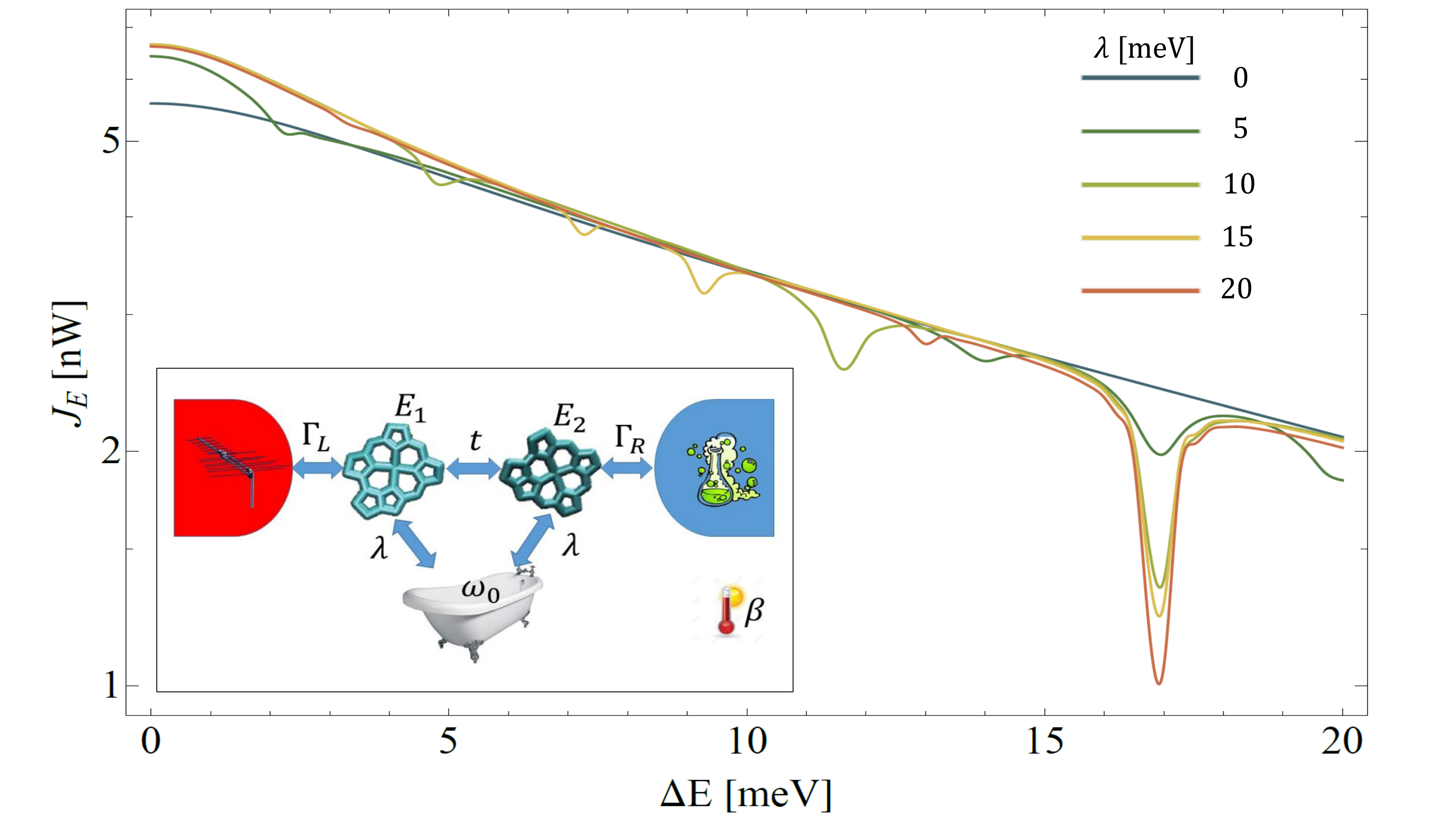}
\caption{Exciton current as a function of $\Delta E$, The energy difference between the two chromophores, for different values of exciton-phonon coupling, $\lambda=0,5,10,15,20$ meV (see text for other numerical parameters). Inset: schematic illustration of the system; a bi-chromophore chain coupled to a source and a sink and global phonon bath.} \label{fig:1}
\end{figure}

%\sec{Exciton transport in the presence of a global phonon}
Our model for exciton transport in the presence of a global phonon (schematically depicted in the inset of Fig~.\ref{fig:1}) is a simple tight-binding model, comprising coupled chromophores on which excitons can reside. One chromophore (e.g. left) is coupled to a source (antenna), and the other (right) is coupled to a sink (reaction center). On the chromophores, the excitons are coupled to a localized vibration.  The system Hamiltonian is thus
\beqa
\mathcal{H} &=&\mathcal{H}_{ex}+\mathcal{H}_{ph}+\cH_{ex-ph} \nonumber \\
\cH_{ex}&=&\sum_{i=1,2} E_i d^\dagger_i d_i -t(d^\dagger_1d_2+h.c.)~~,
\eeqa\label{Ham1}
where $d^\dagger_i (d_i)$ creates (annihilates) an exciton on chromophore $i=1,2$, $t$ is the inter-chromophore excitonic hopping amplitude, and $E_i$ are the exciton energies. A central parameter is the exciton energy difference, $\Delta E=E_1-E_2$. The phonon and the exciton-phonon coupling Hamiltonians take the respective forms 
 \beqa  
 \cH_{ph}+\cH_{ex-ph}&=& \hbar \omega_0 a^\dagger a+\lambda (a+a^\dagger)(d^\dagger_1 d_1+ d^\dagger_2 d_2)~~,\nonumber\\ \eeqa \label{H:ex-ph} 
 where $a^\dagger (a)$ creates (annihilates) a phonon with frequency $\omega_0$, and $\lambda$ is the exciton-phonon coupling. In general, the couplings between the excitons and the vibration (even for the global phonon case) need not be equal, a situation which would lead to an effective reorganization energy for exciton transfer. However, this situation does not allow for an analytic solution as we present below, and is therefore left for future studies. 
 
%In the "local phonon" case, the two operators $a$ and $b$ commute with each other, representing independent bosons. In the "global phonon" case, they represent the same vibration, i.e. obey the bosonic commutation relations, $[a^\dagger,b]=-1$. The Hamiltonian of Eq.\ref{H:ex-ph} can be transformed into a global-phonon form with the substitution $b \rightarrow a,\omega\rightarrow \omega/2, \lambda \rightarrow \lambda /2$.

%To this Hamiltonian description we add several ingredients. First, we assume that the phonons are coupled to a thermal bath, with a temperature determined externally, i.e. it is a parameter of the system. Second, we assume that the excitonic system is in contact with reservoirs that push excitons into the left site and extract them from the right. These excitonic reservoirs drive the system out of equilibrium, and therefore there is excitonic current through the system. This current is a measure of the junction's transfer efficiency, and is almost proportional to the energy current flowing through the system [SEE LATER]

The Hamiltonian of Eq.~\ref{Ham1}-\ref{H:ex-ph} (in the absence of reservoirs) can be diagonalized exactly, by first diagonalizing the exciton part and then performing a polaron transformation \cite{Ren2012,Wang2012a,Chen2008} (see SI for details). The resulting eigenvalues are polaron states, 
defined by the polaron quantum numbers $n_p^{(0)},n_p^{(+)},n_p^{(-)},n_p^{(2)}$, corresponding to the number of polarons in the subspaces of no excitons (0), single exciton in the bonding (+) and anti-bonding (-) states, and two excitons (2),
%\begin{widetext}
\beqa \label{eigencase1}
	&\cH \ket{n_+=0, n_-=0,n^{(0)}_{p}} =\omega_0 n_{p}\ket{n_+=0, n_-=0, n^{(0)}_{p}} \equiv \epsilon_0(n_p) \ket{0,0,n^{(0)}_{p}} ,
	\\ \nonumber
	&\cH \ket{1,0,n^{(+)}_{p}} =\big(E_+ +\omega_0 n_{p} -\frac{\lambda^2}{\omega_0}\big)\ket{1,0,  n^{(+)}_{p}} \equiv
	\epsilon_+ (n_p) \ket{1,0, n^{(+)}_{p}} ,
	\\ \nonumber
	&\cH \ket{0, 1, n^{(-)}_{p}} =\big(E_- +\omega_0 n_{p} -\frac{\lambda^2}{\omega_0}\big)\ket{0, 1, n^{(-)}_{p}}
	\equiv \epsilon_- (n_p) \ket{0,1,n^{(-)}_{p}} ,
	\\ \nonumber
	&\cH \ket{1, 1, n^{(2)}_{p}} =\big(2\bar{E}+ \omega_0 n_{p} -\frac{4\lambda^2}{\omega_0}\big) \ket{1, 1, n^{(2)}_{p}} \equiv \epsilon_2(n_p) \ket{1,1,n^{(2)}_{p}} ~~.
\eeqa
%\end{widetext}
The energies $E_\pm$ are the eigenvalues of the exciton Hamiltonian $E_\pm=\bar{E}\pm\sqrt{\Delta E^2+4t^2}/2$, and $\bar{E}\equiv(E_1+E_2)/2$. 

Next, we evaluate the single-exciton bare (i.e. in the absence of the reservoirs) retarded and advances Green's functions (which will be required for calculating the current), $g^{r(a)}(\e)$,  which are $2\times2$ matrices in chromophore space. 
We employ the Lehman representation \cite{mahan2013many},
\beq \label{readGF}
g_{ij}^{r(a)}(\e) = -\frac{1}{\cZ}\sum_{\phi,\psi} \frac{e^{-\beta E_\psi}+e^{-\beta E_\phi}}{E_\phi-E_\psi-\e\pm i\eta} \langle \psi  | d^\dagger_i |\phi \rangle \langle \phi | d_j | \psi \rangle~~,  
\eeq 
where $i,j$ are site (chromophore) indices, $|\phi \rangle,|\psi\rangle$ are the eigenvectors of the Hamiltonian with the corresponding eigen-energies $E_\phi, E_\psi$, and $\cZ=\sum_\phi e^{-\beta E_\phi}$ is the partition function, where $k_B T=\beta^{-1}$ is the temperature of the system. Plugging into Eq.~\ref{readGF} the eigenvalues of Eq.~\ref{eigencase1} one obtains the bare Green's functions. For instance, the off-diagonal Green's functions are given by  
 %\begin{widetext}
 \beqa \label{retgre}
			g_{12}^{r}(\xi) &=&-\frac{1}{\cZ }\frac{t}{\sqrt{\Delta E^2+4t^2}}  \left[ 
		\sum_{n_p,m_p}  \left|D_{n_p,m_p}(\lambda,\omega_0) \right|^2  \frac{e^{-\beta \epsilon_{+}(n_p)}+e^{-\beta \epsilon_{0}(m_p)}}{ \epsilon_{+}(n_p)-\epsilon_{0}(m_p)-\xi + i 0^+} \right.
		\nonumber \\ &-&\sum_{n_p,m_p}  \left|D_{n_p,m_p}(\lambda,\omega_0) \right|^2  \frac{e^{-\beta \epsilon_{-}(n_p)}+e^{-\beta \epsilon_{0}(m_p)}}{ \epsilon_{-}(n_p)-\epsilon_{0}(m_p)-\xi + i 0^+}  \nonumber \\
		  &-&\sum_{n_p,m_p}  \left|D_{n_p,m_p}(\lambda,\omega_0) \right|^2  \frac{e^{-\beta \epsilon_{2}(n_p)}+e^{-\beta \epsilon_{+}(m_p)}}{ \epsilon_{2}(n_p)-\epsilon_{+}(m_p)-\xi + i 0^+} \nonumber\\ 
	  &+&\sum_{n_p,m_p}\left. \left|D_{n_p,m_p}(\lambda,\omega_0) \right|^2  \frac{e^{-\beta \epsilon_{2}(n_p)}+e^{-\beta \epsilon_{-}(m_p)}}{ \epsilon_{2}(n_p)-\epsilon_{-}(m_p)-\xi + i 0^+} \right] 	 ~~  ,
\eeqa 
where $D_{n_p,m_p} (\lambda,\omega_0))$ are the Franck-Condon factors of polaron-polaron overlap (see SI for details). %$\cZ$ is the partition function and $\beta=(k_B T)^{-1}$ is the inverse temperature.

Up until now, the exciton-phonon system was isolated from the reservoirs (and thus at thermal equilibrium), and the expression for $g_{ij}$ was exact. We proceed to incorporate the source and sink into the formulation. Using the bare Green's functions, the full Green's functions are evaluated from $
G^{r(a)}\approx\left( \left(g^{r(a)}\right)^{-1}+\Sigma^{r(a)}\right)^{-1} $ , where $\Sigma^{r(a)}(\e)$ are the self energies describing the hopping between the chromophors and the source and sink reservoirs. This equation for the full Green's function is an approximation (usually coined the non-crossing approximation) and is applicable for small system-reservoir couplings, as is the case here \cite{Ren2012}. For simplicity and following Ref.~\citenum{Ren2012} we assume an energy-independent self-energies (wide-band approximation), with the source and sink terms coupling only to the left and right chromophores, respectively, leading to the form for the self energies, $\Sigma^{r(a)}=\Sigma^{r(a)}_L+\Sigma^{r(a)}_R$, where 

 $\Sigma^{r(a)}_L=\mp i \begin{pmatrix} \Gamma_L & 0 \\ 0 & 0 \end{pmatrix}$, $\Sigma^{r(a)}_R=\mp i \begin{pmatrix} 0 & 0 \\ 0 & \Gamma_R \end{pmatrix}$ and $\Gamma_{L,R}$ are the broadenings (inverse life-times) due to the hopping to the source  and sink, respectively, taken for simplicity to be equal $ \Gamma_L= \Gamma_R= \Gamma$. From the full Green's function, the exciton current is calculated using the Landauer formula \cite{DiVentra2008} $I=\int d \e \cT(\e) \left(f_L(\e)-f_R(\e)\right)$, where $\cT(\e)=\mathrm{Tr} \left(\Sigma^r_L G^r(\e) \Sigma^a_R G^a(\e)\right) $ \cite{Meir1992}. The distributions $f_L,f_R$ describe the probability of finding an exciton with energy $\e$ at the source and sink, respectively, in similarity to electron transport (where they are simply the Fermi functions of the corresponding electrodes). Here we follow Ref.~\citenum{Pelzer2014} and take $f_L=1,f_R=0$, implying that there is no "backflow" of excitons from the sink to the source, a situation which is similar to an electronic molecular junction at high voltage. We note that in our case, due to the large exciton energy, $\bar{E}>>\Delta E$, the exciton and energy currents are almost exactly proportional to each other (see SI for details).

Fig.~\ref{fig:1} shows the exciton current as a function of the chromophore energy difference $\Delta E$, for different values of the exciton-phonon couplings, $\lambda=0,5,10,15,20$ meV. Other numerical parameters are $\Gamma=1.45$ meV, $t=1$ meV, $\omega_0=17$ meV, and $\beta=0.039$ meV$^{-1}$ (corresponding to room temperature). These values are taken to agree with the presumed values of biological Exciton transfer complexes \cite{Cho2005}.  

\begin{figure}[h!]
\vskip 0.5cm
\includegraphics[width=3.5 in]{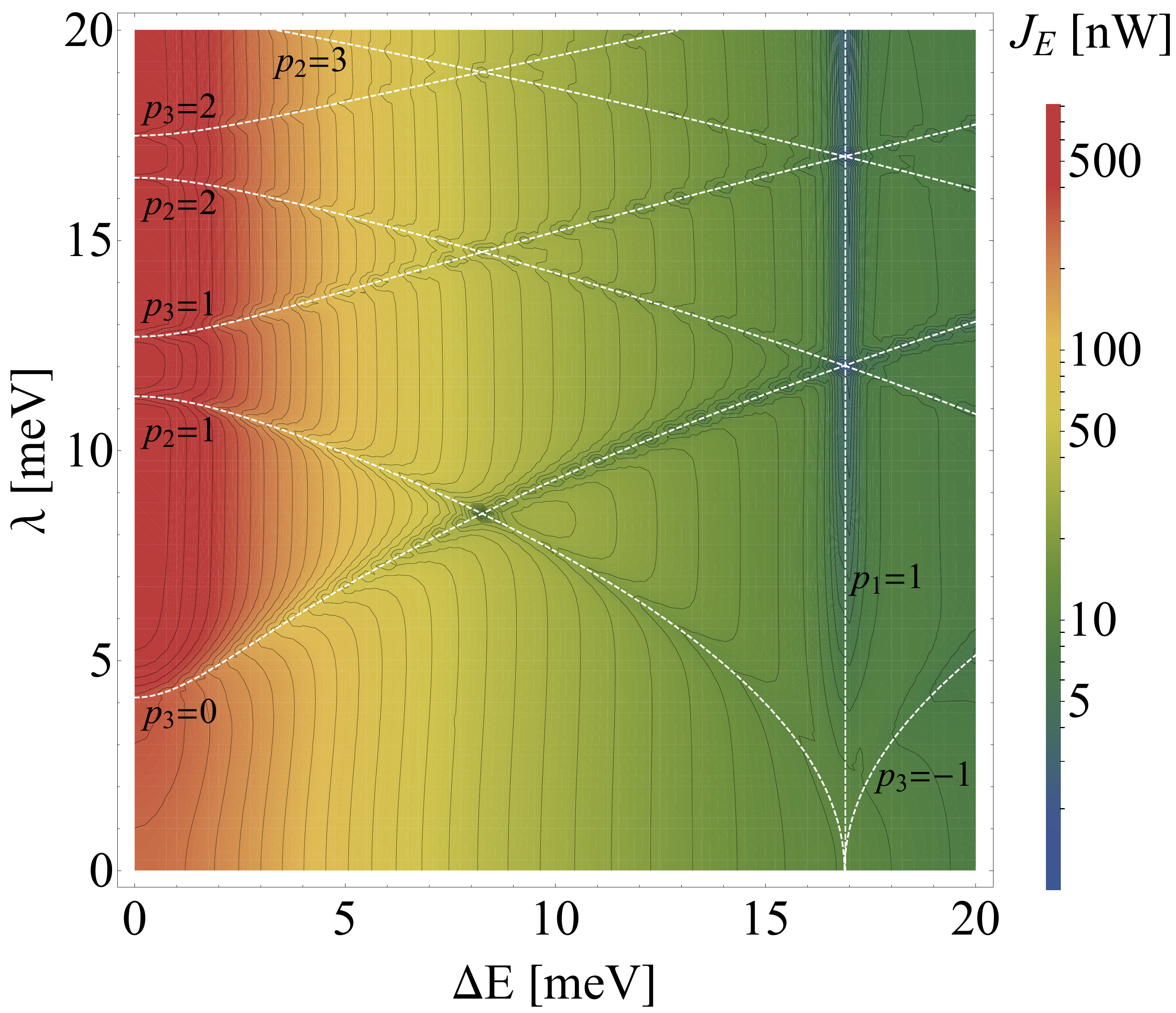}
\caption{Color-plotted exciton current as a function of $\Delta E$ and $\lambda$, exhibiting quantum anti-resonances. Dashed lines correspond to the solutions of the analytic anti-resonance conditions of Eq.~\ref{ARconditions}, pointing at quantum interference as the source of the ARs. } \label{fig:2}
\end{figure}

The most striking feature in Fig.~\ref{fig:1} is the appearance of anti-resonances (ARs), namely sharp dips in the current. To understand the origin of these ARs, we turn back to the formula for the current. One can easily check (see supplementary information) that the transmission function (related to the total current via integration) is proportional to $|g^r_{12}|^2$, which implies that reduction of current could be related to the properties of the bare Green's function. Indeed, if one looks at the bare Green's function (Eq.~\ref{retgre}), one can see that it is composed of four terms with alternating signs. These correspond to different processes in which an exciton can be added or removed from the system. The ARs occur when pairs of these processes cancel each other - which is nothing but destructive quantum interference between the different transfer paths. For this to happen, the denominators of the different terms must be equal, which occur at the following conditions, 
\beqa \label{ARconditions}
&\epsilon_--\epsilon_+=\omega_0 p_1 \\ \nonumber
&\epsilon_2+\epsilon_0-2\epsilon_+=\omega_0 p_2 \\ \nonumber
&\epsilon_2+\epsilon_0-2\epsilon_-=\omega_0 p_3~~,
\eeqa 
where $p_i$ are arbitrary integers. These conditions can be solved for $\Delta E$ and $\lambda$, to generate the AR lines. In Fig.~\ref{fig:2} the current is color-plotted as a function of $\Delta E$ and $\lambda$ (other parameters are same as in Fig.~\ref{fig:1}). The dashed lines correspond to Eqs.~\ref{ARconditions}, and clearly coincide with the numerically evaluated ARs. 

The physical interpretation of the ARs is as follows. As an exciton is injected into the system from the source, it's wave function is composed of a superposition of bonding (+) and anti-bonding (-) states, thermally weighted with different polaron numbers. When the AR conditions are satisfied, the different parts of the wave function %(characterized by occupying either 2 excitons, the bonding or anti-bonding state or zero excitons, and with polaron numbers differing by an integer $p$)  
propagate in-phase from the left chromophore (coupled to the source) to the right chromophore (coupled to the sink), leading to destructive quantum interference. This is a unique situation in which the exciton-polaron states interfere with each other. Interestingly, there is no constructive  interference, since the conditions for it are never satisfied (see supplementary material). We therefore conclude that, within a single-vibration model, the global phonon only hampers exciton transfer, as it generates anti-resonances on top of the bare current curve.  

\begin{figure}[h!]
\vskip 0.5cm
\includegraphics[width=3.5 in]{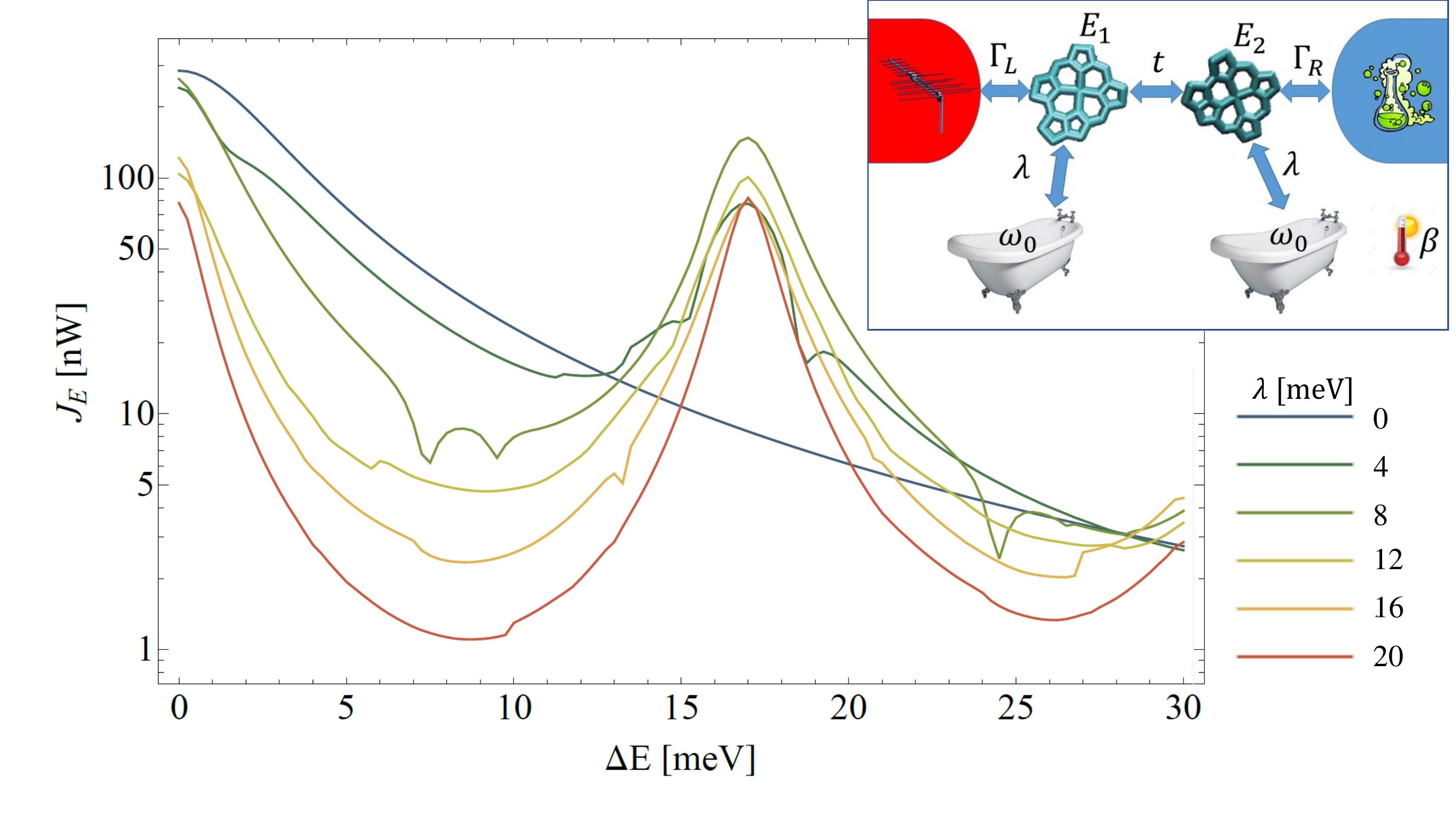}
\caption{Local phonon case: current as a function of $\Delta E$ for different values of the electron phonon coupling, $\lambda=0,4,...20$ meV (other parameters are the same as in Fig.~\ref{fig:1}).  } \label{fig:3}
\end{figure}

%\sec{Exciton transport in the presence of a local phonon} 
The second case we consider is the case in which each chromophore is coupled to a separate vibration, schematically depicted in the inset of Fig.~\ref{fig:3} (which seems to be the case in natural systems \cite{maly2016, Aghtar2014, tiwari2013}). The system is then described by the Hamiltonian
\beqa
\mathcal{H} &=&\mathcal{H}_{ex}+\mathcal{H}_{ph}+\cH_{ex-ph} \nonumber \\
\cH_{ex}&=&\sum_{i=1,2} E_i d^\dagger_i d_i -t(d^\dagger_1d_2+h.c.)\nonumber \\
\cH_{ph}+\cH_{ex-ph}&=& \nonumber \\
 \sum_{i=1,2}&&\left( \hbar \omega_0  a^\dagger_i a_i + \lambda (a_i+a^\dagger_i)d^\dagger_i d_i\right)~~,
\eeqa\label{Ham2}
 where $a^\dagger_i (a_i)$ creates (annihilates) a phonon at the chromophore $i=1,2$. 
As opposed to the global-phonon case, in this situation the Hamiltonian cannot be diagonalized exactly. To proceed, we start by performing a local polaron transformation (see supplementary information), then re-writing the exciton hopping term in the polaron basis. This allows for an efficient numerical diagonalization of the Hamiltonian and evaluation of the Green's function, which are then used to calculate the transmission function and the current. 

In Fig.~\ref{fig:3} The current is plotted as a function of $\Delta E$ for different values of the electron-phonon coupling, $\lambda=0,4,...20$ meV (other parameters are the same as in Fig.~\ref{fig:1}).  Comparing to the local-phonon case of Fig.~\ref{fig:1}, we find substantial qualitative differences. Now, the current is characterized by broad resonances which appear when $\Delta E$ is integer multiples of $\omega_0$. This is a simple phonon-assisted tunneling resonance, occurring whenever the transmission peak from the left chromophore (which is broadened by the coupling to the reservoirs and shifted by the local vibration) aligns with the transmission peak from the right  chromophore (which is also broadened and shifted) \cite{Dijkstra2015,Novoderezhkin2017}.

Vestiges of anti-resonances can still be seen even for the local phonon case for small values of $\lambda$. This can be explained by noting that the exciton tunneling term, when translated into the polaron basis, couples states at chromophore $1$ with any polaron number $n_p$ to states at chromophore $2$ with any other polaron number $m_p$. However, when $\lambda<<\omega_0$ the overlap integrals are such that, effectively, only polarons with the same polaron number are coupled by the exciton hopping term. Thus, for $\lambda<<\omega_0$, the local phonon Hamiltonian resembles the global-phonon one, and exhibits similar behavior, including the anti-resonances. 

\begin{figure}[h!]
\vskip 0.5cm
\includegraphics[width=3.5 in]{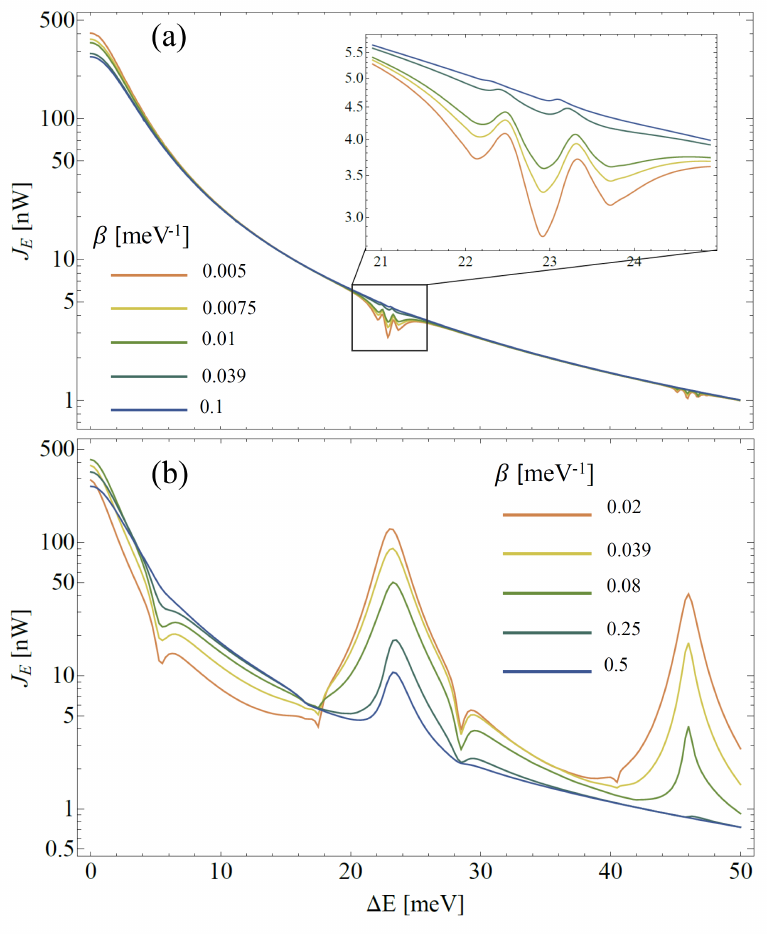}%{fig4hrzntl.pdf}%
\caption{Current as a function of $\Delta E$ for different inverse temperatures $\beta$, for the global phonon (a) and local phonon (b) cases. Inriguingly, the features become more distinct with increasing temperature.} \label{fig:4}
\end{figure}

Finally, we turn to the temperature dependence of the current. As was demonstrated above, the current (as a function of $\Delta E$) is characterized by features which are of quantum-mechanical origin, namely destructive interference (for the global-phonon case) and resonant tunneling (for the local phonon case). Naively, one expects that any quantum-mechanical signatures would be washed out by increasing temperature. Surprisingly, this is not the case here. 
In Fig.~\ref{fig:4} the current as a function of $\Delta E$ is plotted for different inverse temperatures $\beta$, for the global phonon (Fig.~\ref{fig:4}(a)) and local phonon (Fig.~\ref{fig:4}(b)) cases. Numerical parameters are $\lambda=8 $ meV,  $\omega_0=23$ meV,  $t=1 $ meV and $\Gamma=1.45$ meV. It can be clearly seen that as the temperatures decrease ($\beta$ increases, marked in the figure from red to blue), the  quantum features in both cases decrease.

To understand this result, consider the global phonon case (where an analytic expression for the Green's function is available, Eq.~\ref{retgre}). As described above, the AR are due to interference - cancellation of different parts of in the Green's function, corresponding to different polaron states, i.e. excitation with different number of excited bosons. Determined by the thermal bath, the weight of these different states is described by a Bolzmann factor. This means that as temperatures go down, the weight of the higher-polaron-number states decreases, and only the ground state becomes important. As a result, high-polaron-number states cannot participate in the interference, and the AR feature are suppressed. Similar argument can be made for the local phonon case, as the resonant tunneling is due to the alignment of transmission peaks separated by the energy of a single vibration (polaron). Since again the excitation of polarons is suppressed with decreasing temperatures, the effect of the resonant tunneling is reduced. We note that our model is limited in treating temperature effects (which appear only through the thermal bath, Eq.~\ref{readGF}). Clearly, as the temperature is increased in natural systems, other effects may come into play: additional vibrational modes might become important, the effective inter-chromophore couplings might change, or correlations between the vibrations might change, all affecting the high-temperature behavior of the system.

In summary, in this paper we calculated the exciton current through a bi-chromophore system in the presence of vibrations, either localized on the chromophores (local phonons) or delocalized over the entire system (global phonons). The two cases exhibit qualitatively different features; global phonons lead to anti-resonances in the current, while local phonons lead to resonant enhancement of the current. This is in line with the observation that in natural photosynthetic exciton transfer systems vibrations are indeed localized around the chromophores, and with the assumption that these systems evolved for optimal exciton transfer. Furthermore, our results show strong dependence of  current on chromophore energies, with large enhancement of the current at specific resonant energy difference values. Such considerations may be important for future design of bio-mimetic artificial light-harvesting systems. 

Finally, we note that our conclusions are relevant also for molecular junctions, since the Hamiltonian description and formalism presented here are equivalent to bi-molecular junctions at high voltages \cite{Sowa2017a,Park2011}, with electric current replacing exciton current. Since measurement of electrical transport is substantially easier than measureing exciton transport (since excitons carry no electrical charge),  molecular junctions are good candidates for verifying our theoretical results. 

\paragraph*{Acknowledgements: } This research was supported in part by the Adelis Foundation for renewable energy research, and by ISF grant 292/15.

\begin{suppinfo} Exact diagonalization of the electron-phonon Hamiltonian. Polaronic Frank-Condon terms. Expressions for exciton and energy currents. Equivalence between exciton and energy currents. Conditions for destructive and constructive interference. 
\end{suppinfo}

%\nocite{*} % Print all references regardless of whether they were cited in the poster or not
%\bibliographystyle{acs} % Plain referencing style
%\bibliography{refs_ExcTrans} 

\begin{mcitethebibliography}{53}
	\providecommand*\natexlab[1]{#1}
	\providecommand*\mciteSetBstSublistMode[1]{}
	\providecommand*\mciteSetBstMaxWidthForm[2]{}
	\providecommand*\mciteBstWouldAddEndPuncttrue
	{\def\EndOfBibitem{\unskip.}}
	\providecommand*\mciteBstWouldAddEndPunctfalse
	{\let\EndOfBibitem\relax}
	\providecommand*\mciteSetBstMidEndSepPunct[3]{}
	\providecommand*\mciteSetBstSublistLabelBeginEnd[3]{}
	\providecommand*\EndOfBibitem{}
	\mciteSetBstSublistMode{f}
	\mciteSetBstMaxWidthForm{subitem}{(\alph{mcitesubitemcount})}
	\mciteSetBstSublistLabelBeginEnd
	{\mcitemaxwidthsubitemform\space}
	{\relax}
	{\relax}
	
	\bibitem[Engel \latin{et~al.}(2007)Engel, Calhoun, Read, Ahn, ManÄal, Cheng,
	Blankenship, and Fleming]{engel2007}
	Engel,~G.~S.; Calhoun,~T.~R.; Read,~E.~L.; Ahn,~T.-K.; ManÄal,~T.;
	Cheng,~Y.-C.; Blankenship,~R.~E.; Fleming,~G.~R. Evidence for Wavelike Energy
	Transfer through Qquantum Coherence in Photosynthetic Systems. \emph{Nature}
	\textbf{2007}, \emph{446}, 782--786\relax
	\mciteBstWouldAddEndPuncttrue
	\mciteSetBstMidEndSepPunct{\mcitedefaultmidpunct}
	{\mcitedefaultendpunct}{\mcitedefaultseppunct}\relax
	\EndOfBibitem
	\bibitem[Calhoun \latin{et~al.}(2009)Calhoun, Ginsberg, Schlau-Cohen, Cheng,
	Ballottari, Bassi, and Fleming]{Calhoun2009}
	Calhoun,~T.~R.; Ginsberg,~N.~S.; Schlau-Cohen,~G.~S.; Cheng,~Y.-C.;
	Ballottari,~M.; Bassi,~R.; Fleming,~G.~R. Quantum Coherence Enabled
	Determination of the Energy Landscape in Light-Harvesting Complex II.
	\emph{The Journal of Physical Chemistry B} \textbf{2009}, \emph{113},
	16291--16295\relax
	\mciteBstWouldAddEndPuncttrue
	\mciteSetBstMidEndSepPunct{\mcitedefaultmidpunct}
	{\mcitedefaultendpunct}{\mcitedefaultseppunct}\relax
	\EndOfBibitem
	\bibitem[Collini \latin{et~al.}(2010)Collini, Wong, Wilk, Curmi, Brumer, and
	Scholes]{Collini2010}
	Collini,~E.; Wong,~C.~Y.; Wilk,~K.~E.; Curmi,~P.~M.; Brumer,~P.; Scholes,~G.~D.
	Coherently Wired Light-Harvesting in Photosynthetic Marine Algae at Ambient
	Temperature. \emph{Nature} \textbf{2010}, \emph{463}, 644--647\relax
	\mciteBstWouldAddEndPuncttrue
	\mciteSetBstMidEndSepPunct{\mcitedefaultmidpunct}
	{\mcitedefaultendpunct}{\mcitedefaultseppunct}\relax
	\EndOfBibitem
	\bibitem[Panitchayangkoon \latin{et~al.}(2010)Panitchayangkoon, Hayes,
	Fransted, Caram, Harel, Wen, Blankenship, and Engel]{Panitchayangkoon2010}
	Panitchayangkoon,~G.; Hayes,~D.; Fransted,~K.~A.; Caram,~J.~R.; Harel,~E.;
	Wen,~J.; Blankenship,~R.~E.; Engel,~G.~S. Long-Lived Quantum Coherence in
	Photosynthetic Complexes at Physiological Temperature. \emph{Proceedings of
		the National Academy of Sciences} \textbf{2010}, \emph{107},
	12766--12770\relax
	\mciteBstWouldAddEndPuncttrue
	\mciteSetBstMidEndSepPunct{\mcitedefaultmidpunct}
	{\mcitedefaultendpunct}{\mcitedefaultseppunct}\relax
	\EndOfBibitem
	\bibitem[Fleming and Scholes(2004)Fleming, and Scholes]{fleming2004}
	Fleming,~G.~R.; Scholes,~G.~D. Physical Chemistry: Quantum Mechanics for
	Plants. \emph{Nature} \textbf{2004}, \emph{431}, 256--257\relax
	\mciteBstWouldAddEndPuncttrue
	\mciteSetBstMidEndSepPunct{\mcitedefaultmidpunct}
	{\mcitedefaultendpunct}{\mcitedefaultseppunct}\relax
	\EndOfBibitem
	\bibitem[Levi \latin{et~al.}(2015)Levi, Mostarda, Rao, and Mintert]{levi2015}
	Levi,~F.; Mostarda,~S.; Rao,~F.; Mintert,~F. Quantum Mechanics of Excitation
	Transport in Photosynthetic Complexes: a key issues review. \emph{Rep. Prog.
		Phys.} \textbf{2015}, \emph{78}, 082001\relax
	\mciteBstWouldAddEndPuncttrue
	\mciteSetBstMidEndSepPunct{\mcitedefaultmidpunct}
	{\mcitedefaultendpunct}{\mcitedefaultseppunct}\relax
	\EndOfBibitem
	\bibitem[Ishizaki and Fleming(2012)Ishizaki, and Fleming]{Ishizaki2012}
	Ishizaki,~A.; Fleming,~G.~R. Quantum Coherence in Photosynthetic Light
	Harvesting. \emph{Annu. Rev. Condens. Matter Phys.} \textbf{2012}, \emph{3},
	333--361\relax
	\mciteBstWouldAddEndPuncttrue
	\mciteSetBstMidEndSepPunct{\mcitedefaultmidpunct}
	{\mcitedefaultendpunct}{\mcitedefaultseppunct}\relax
	\EndOfBibitem
	\bibitem[Collini(2013)]{Collini2013}
	Collini,~E. Spectroscopic Signatures of Quantum-Coherent Energy Transfer.
	\emph{Chemical Society Reviews} \textbf{2013}, \emph{42}, 4932--4947\relax
	\mciteBstWouldAddEndPuncttrue
	\mciteSetBstMidEndSepPunct{\mcitedefaultmidpunct}
	{\mcitedefaultendpunct}{\mcitedefaultseppunct}\relax
	\EndOfBibitem
	\bibitem[Lambert \latin{et~al.}(2013)Lambert, Chen, Cheng, Li, Chen, and
	Nori]{Lambert2013}
	Lambert,~N.; Chen,~Y.-N.; Cheng,~Y.-C.; Li,~C.-M.; Chen,~G.-Y.; Nori,~F.
	Quantum Biology. \emph{Nature Physics} \textbf{2013}, \emph{9}, 10--18\relax
	\mciteBstWouldAddEndPuncttrue
	\mciteSetBstMidEndSepPunct{\mcitedefaultmidpunct}
	{\mcitedefaultendpunct}{\mcitedefaultseppunct}\relax
	\EndOfBibitem
	\bibitem[Pach{\'o}n and Brumer(2012)Pach{\'o}n, and Brumer]{Pachon2012}
	Pach{\'o}n,~L.~A.; Brumer,~P. Computational Methodologies and Physical Insights
	into Electronic Energy Transfer in Photosynthetic Light-Harvesting Complexes.
	\emph{Physical Chemistry Chemical Physics} \textbf{2012}, \emph{14},
	10094--10108\relax
	\mciteBstWouldAddEndPuncttrue
	\mciteSetBstMidEndSepPunct{\mcitedefaultmidpunct}
	{\mcitedefaultendpunct}{\mcitedefaultseppunct}\relax
	\EndOfBibitem
	\bibitem[Kassal \latin{et~al.}(2013)Kassal, Yuen-Zhou, and
	Rahimi-Keshari]{kassal2013}
	Kassal,~I.; Yuen-Zhou,~J.; Rahimi-Keshari,~S. Does Coherence Enhance Transport
	in Photosynthesis. \emph{J. Phys. Chem. Lett.} \textbf{2013}, \emph{4},
	362--367\relax
	\mciteBstWouldAddEndPuncttrue
	\mciteSetBstMidEndSepPunct{\mcitedefaultmidpunct}
	{\mcitedefaultendpunct}{\mcitedefaultseppunct}\relax
	\EndOfBibitem
	\bibitem[Miller(2012)]{Miller2012}
	Miller,~W.~H. Perspective: Quantum or Classical Coherence? \emph{The Journal of
		chemical physics} \textbf{2012}, \emph{136}, 210901\relax
	\mciteBstWouldAddEndPuncttrue
	\mciteSetBstMidEndSepPunct{\mcitedefaultmidpunct}
	{\mcitedefaultendpunct}{\mcitedefaultseppunct}\relax
	\EndOfBibitem
	\bibitem[Ritschel \latin{et~al.}(2011)Ritschel, Roden, Strunz, Aspuru-Guzik,
	and Eisfeld]{ritschel2011}
	Ritschel,~G.; Roden,~J.; Strunz,~W.~T.; Aspuru-Guzik,~A.; Eisfeld,~A. Absence
	of Quantum Oscillations and Dependence on Site Energies in Electronic
	Excitation Transfer in the Fenna-Matthews-Olson Trimer. \emph{J. Phys. Chem.
		Lett.} \textbf{2011}, \emph{2}, 2912--2917\relax
	\mciteBstWouldAddEndPuncttrue
	\mciteSetBstMidEndSepPunct{\mcitedefaultmidpunct}
	{\mcitedefaultendpunct}{\mcitedefaultseppunct}\relax
	\EndOfBibitem
	\bibitem[Duan \latin{et~al.}(2017)Duan, Prokhorenko, Cogdell, Ashraf, Stevens,
	Thorwart, and Miller]{duan2017nature}
	Duan,~H.-G.; Prokhorenko,~V.~I.; Cogdell,~R.~J.; Ashraf,~K.; Stevens,~A.~L.;
	Thorwart,~M.; Miller,~R.~D. Nature Does Not Rely on Long-Lived Electronic
	Quantum Coherence for Photosynthetic Energy Transfer. \emph{Proceedings of
		the National Academy of Sciences} \textbf{2017}, \emph{114}, 8493--8498\relax
	\mciteBstWouldAddEndPuncttrue
	\mciteSetBstMidEndSepPunct{\mcitedefaultmidpunct}
	{\mcitedefaultendpunct}{\mcitedefaultseppunct}\relax
	\EndOfBibitem
	\bibitem[Wilkins and Dattani(2015)Wilkins, and Dattani]{Wilkins2015}
	Wilkins,~D.~M.; Dattani,~N.~S. Why Quantum Coherence is Not Important in the
	Fenna--Matthews--Olsen Complex. \emph{Journal of chemical theory and
		computation} \textbf{2015}, \emph{11}, 3411--3419\relax
	\mciteBstWouldAddEndPuncttrue
	\mciteSetBstMidEndSepPunct{\mcitedefaultmidpunct}
	{\mcitedefaultendpunct}{\mcitedefaultseppunct}\relax
	\EndOfBibitem
	\bibitem[Mohseni \latin{et~al.}(2008)Mohseni, Rebentrost, Lloyd, and
	Aspuru-Guzik]{mohseni2008}
	Mohseni,~M.; Rebentrost,~P.; Lloyd,~S.; Aspuru-Guzik,~A. Environment-Assisted
	Quantum Walks in Photosynthetic Energy Transfer. \emph{Journal of Chemical
		Physics} \textbf{2008}, \emph{129}, 174106\relax
	\mciteBstWouldAddEndPuncttrue
	\mciteSetBstMidEndSepPunct{\mcitedefaultmidpunct}
	{\mcitedefaultendpunct}{\mcitedefaultseppunct}\relax
	\EndOfBibitem
	\bibitem[Plenio and Huelga(2008)Plenio, and Huelga]{plenio2008}
	Plenio,~M.~B.; Huelga,~S.~F. Dephasing-Assisted Transport: Quantum Networks and
	Biomolecules. \emph{New J. Phys.} \textbf{2008}, \emph{10}, 113019\relax
	\mciteBstWouldAddEndPuncttrue
	\mciteSetBstMidEndSepPunct{\mcitedefaultmidpunct}
	{\mcitedefaultendpunct}{\mcitedefaultseppunct}\relax
	\EndOfBibitem
	\bibitem[Rebentrost \latin{et~al.}(2009)Rebentrost, Mohseni, Kassal, Lloyd, and
	Aspuru-Guzik]{Rebentrost2009}
	Rebentrost,~P.; Mohseni,~M.; Kassal,~I.; Lloyd,~S.; Aspuru-Guzik,~A.
	Environment-Assisted Quantum Transport. \emph{New J. Phys.} \textbf{2009},
	\emph{11}, 033003\relax
	\mciteBstWouldAddEndPuncttrue
	\mciteSetBstMidEndSepPunct{\mcitedefaultmidpunct}
	{\mcitedefaultendpunct}{\mcitedefaultseppunct}\relax
	\EndOfBibitem
	\bibitem[Chin \latin{et~al.}(2010)Chin, Datta, Caruso, Huelga, and
	Plenio]{chin2010}
	Chin,~A.~W.; Datta,~A.; Caruso,~F.; Huelga,~S.~F.; Plenio,~M.~B. Noise-Assisted
	Energy Transfer in Quantum Networks and Light-Harvesting Complexes. \emph{New
		J. Phys.} \textbf{2010}, \emph{12}, 065002\relax
	\mciteBstWouldAddEndPuncttrue
	\mciteSetBstMidEndSepPunct{\mcitedefaultmidpunct}
	{\mcitedefaultendpunct}{\mcitedefaultseppunct}\relax
	\EndOfBibitem
	\bibitem[Wu \latin{et~al.}(2010)Wu, Liu, Shen, Cao, and Silbey]{Wu2010}
	Wu,~J.; Liu,~F.; Shen,~Y.; Cao,~J.; Silbey,~R.~J. Efficient Energy Transfer in
	Light-Harvesting Systems, I: Optimal Temperature, Reorganization Energy and
	Spatial--Temporal Correlations. \emph{New Journal of Physics} \textbf{2010},
	\emph{12}, 105012\relax
	\mciteBstWouldAddEndPuncttrue
	\mciteSetBstMidEndSepPunct{\mcitedefaultmidpunct}
	{\mcitedefaultendpunct}{\mcitedefaultseppunct}\relax
	\EndOfBibitem
	\bibitem[Wu \latin{et~al.}(2012)Wu, Liu, Ma, Silbey, and Cao]{Wu2012}
	Wu,~J.; Liu,~F.; Ma,~J.; Silbey,~R.~J.; Cao,~J. Efficient Energy Transfer in
	Light-Harvesting Systems: Quantum-Classical Comparison, Flux Network, and
	Robustness Analysis. \emph{The Journal of chemical physics} \textbf{2012},
	\emph{137}, 174111\relax
	\mciteBstWouldAddEndPuncttrue
	\mciteSetBstMidEndSepPunct{\mcitedefaultmidpunct}
	{\mcitedefaultendpunct}{\mcitedefaultseppunct}\relax
	\EndOfBibitem
	\bibitem[Moix \latin{et~al.}(2011)Moix, Wu, Huo, Coker, and Cao]{Moix2011}
	Moix,~J.; Wu,~J.; Huo,~P.; Coker,~D.; Cao,~J. Efficient Energy Transfer in
	Light-Larvesting Systems, III: The Influence of the Eighth
	Bacteriochlorophyll on the Dynamics and Efficiency in FMO. \emph{The Journal
		of Physical Chemistry Letters} \textbf{2011}, \emph{2}, 3045--3052\relax
	\mciteBstWouldAddEndPuncttrue
	\mciteSetBstMidEndSepPunct{\mcitedefaultmidpunct}
	{\mcitedefaultendpunct}{\mcitedefaultseppunct}\relax
	\EndOfBibitem
	\bibitem[Semiao \latin{et~al.}(2010)Semiao, Furuya, and Milburn]{Semiao2010}
	Semiao,~F.~L.; Furuya,~K.; Milburn,~G.~J. Vibration-Enhanced Quantum Transport.
	\emph{New Journal of Physics} \textbf{2010}, \emph{12}, 083033\relax
	\mciteBstWouldAddEndPuncttrue
	\mciteSetBstMidEndSepPunct{\mcitedefaultmidpunct}
	{\mcitedefaultendpunct}{\mcitedefaultseppunct}\relax
	\EndOfBibitem
	\bibitem[Nalbach \latin{et~al.}(2010)Nalbach, Eckel, and Thorwart]{Nalbach2010}
	Nalbach,~P.; Eckel,~J.; Thorwart,~M. Quantum Coherent Biomolecular Energy
	Transfer with Spatially Correlated Fluctuations. \emph{New Journal of
		Physics} \textbf{2010}, \emph{12}, 065043\relax
	\mciteBstWouldAddEndPuncttrue
	\mciteSetBstMidEndSepPunct{\mcitedefaultmidpunct}
	{\mcitedefaultendpunct}{\mcitedefaultseppunct}\relax
	\EndOfBibitem
	\bibitem[Scholak \latin{et~al.}(2011)Scholak, de~Melo, Wellens, Mintert, and
	Buchleitner]{Scholak2011a}
	Scholak,~T.; de~Melo,~F.; Wellens,~T.; Mintert,~F.; Buchleitner,~A. Efficient
	and Coherent Excitation Transfer across Disordered Molecular Networks.
	\emph{Phys. Rev. E} \textbf{2011}, \emph{83}, 021912\relax
	\mciteBstWouldAddEndPuncttrue
	\mciteSetBstMidEndSepPunct{\mcitedefaultmidpunct}
	{\mcitedefaultendpunct}{\mcitedefaultseppunct}\relax
	\EndOfBibitem
	\bibitem[Ajisaka \latin{et~al.}(2015)Ajisaka, Zunkovic, and Dubi]{Ajisaka2015}
	Ajisaka,~S.; Zunkovic,~B.; Dubi,~Y. The Molecular Photo-Cell: Quantum Transport
	and Energy Conversion at Strong Non-Equilibrium. \emph{Sci. Rep.}
	\textbf{2015}, \emph{5}, 8312\relax
	\mciteBstWouldAddEndPuncttrue
	\mciteSetBstMidEndSepPunct{\mcitedefaultmidpunct}
	{\mcitedefaultendpunct}{\mcitedefaultseppunct}\relax
	\EndOfBibitem
	\bibitem[Lim \latin{et~al.}(2014)Lim, Tame, Yee, Lee, and Lee]{Lim2014}
	Lim,~J.; Tame,~M.; Yee,~K.~H.; Lee,~J.-S.; Lee,~J. Phonon-Induced Dynamic
	Resonance Energy Transfer. \emph{New Journal of Physics} \textbf{2014},
	\emph{16}, 053018\relax
	\mciteBstWouldAddEndPuncttrue
	\mciteSetBstMidEndSepPunct{\mcitedefaultmidpunct}
	{\mcitedefaultendpunct}{\mcitedefaultseppunct}\relax
	\EndOfBibitem
	\bibitem[Le{\'o}n-Montiel \latin{et~al.}(2015)Le{\'o}n-Montiel,
	Quiroz-Ju{\'a}rez, Quintero-Torres, Dom{\'\i}nguez-Ju{\'a}rez, Moya-Cessa,
	Torres, and Arag{\'o}n]{leon2015noise}
	Le{\'o}n-Montiel,~R. d.~J.; Quiroz-Ju{\'a}rez,~M.~A.; Quintero-Torres,~R.;
	Dom{\'\i}nguez-Ju{\'a}rez,~J.~L.; Moya-Cessa,~H.~M.; Torres,~J.~P.;
	Arag{\'o}n,~J.~L. Noise-Assisted Energy Transport in Electrical Oscillator
	Networks with Off-Diagonal Dynamical Disorder. \emph{Scientific reports}
	\textbf{2015}, \emph{5}, 17339\relax
	\mciteBstWouldAddEndPuncttrue
	\mciteSetBstMidEndSepPunct{\mcitedefaultmidpunct}
	{\mcitedefaultendpunct}{\mcitedefaultseppunct}\relax
	\EndOfBibitem
	\bibitem[Biggerstaff \latin{et~al.}(2016)Biggerstaff, Heilmann, Zecevik,
	Gr{\"a}fe, Broome, Fedrizzi, Nolte, Szameit, White, and
	Kassal]{Biggerstaff2016}
	Biggerstaff,~D.~N.; Heilmann,~R.; Zecevik,~A.~A.; Gr{\"a}fe,~M.; Broome,~M.~A.;
	Fedrizzi,~A.; Nolte,~S.; Szameit,~A.; White,~A.~G.; Kassal,~I. Enhancing
	Coherent Transport in a Photonic Network Using Controllable Decoherence.
	\emph{Nature communications} \textbf{2016}, \emph{7}, 11282\relax
	\mciteBstWouldAddEndPuncttrue
	\mciteSetBstMidEndSepPunct{\mcitedefaultmidpunct}
	{\mcitedefaultendpunct}{\mcitedefaultseppunct}\relax
	\EndOfBibitem
	\bibitem[Viciani \latin{et~al.}(2016)Viciani, Gherardini, Lima, Bellini, and
	Caruso]{viciani2016disorder}
	Viciani,~S.; Gherardini,~S.; Lima,~M.; Bellini,~M.; Caruso,~F. Disorder and
	Dephasing as Control Knobs for Light Transport in Optical Fiber Cavity
	Networks. \emph{Scientific reports} \textbf{2016}, \emph{6}, 37791\relax
	\mciteBstWouldAddEndPuncttrue
	\mciteSetBstMidEndSepPunct{\mcitedefaultmidpunct}
	{\mcitedefaultendpunct}{\mcitedefaultseppunct}\relax
	\EndOfBibitem
	\bibitem[Caruso \latin{et~al.}(2016)Caruso, Crespi, Ciriolo, Sciarrino, and
	Osellame]{Caruso2016}
	Caruso,~F.; Crespi,~A.; Ciriolo,~A.~G.; Sciarrino,~F.; Osellame,~R. Fast Escape
	of a Quantum Walker from an Integrated Photonic Maze. \emph{Nature
		Communications} \textbf{2016}, \emph{7}, 11682\relax
	\mciteBstWouldAddEndPuncttrue
	\mciteSetBstMidEndSepPunct{\mcitedefaultmidpunct}
	{\mcitedefaultendpunct}{\mcitedefaultseppunct}\relax
	\EndOfBibitem
	\bibitem[Sowa \latin{et~al.}(2017)Sowa, Mol, Briggs, and Gauger]{sowa2017}
	Sowa,~J.~K.; Mol,~J.~A.; Briggs,~G. A.~D.; Gauger,~E.~M. Vibrational Effects in
	Charge Transport through a Molecular Double Quantum Dot. \emph{Physical
		Review B} \textbf{2017}, \emph{95}, 085423\relax
	\mciteBstWouldAddEndPuncttrue
	\mciteSetBstMidEndSepPunct{\mcitedefaultmidpunct}
	{\mcitedefaultendpunct}{\mcitedefaultseppunct}\relax
	\EndOfBibitem
	\bibitem[Sowa \latin{et~al.}(2017)Sowa, Mol, Briggs, and Gauger]{Sowa2017a}
	Sowa,~J.~K.; Mol,~J.~A.; Briggs,~G. A.~D.; Gauger,~E.~M. Environment-Assisted
	Quantum Transport Through Single-Molecule Junctions. \emph{Physical Chemistry
		Chemical Physics} \textbf{2017}, \emph{19}, 29534--29539\relax
	\mciteBstWouldAddEndPuncttrue
	\mciteSetBstMidEndSepPunct{\mcitedefaultmidpunct}
	{\mcitedefaultendpunct}{\mcitedefaultseppunct}\relax
	\EndOfBibitem
	\bibitem[Zerah-Harush and Dubi(2018)Zerah-Harush, and Dubi]{Zerah-Harush2018}
	Zerah-Harush,~E.; Dubi,~Y. Universal Origin for Environment-Assisted Quantum
	Transport in Exciton Transfer Networks. \emph{The journal of physical
		chemistry letters} \textbf{2018}, \emph{9}, 1689\relax
	\mciteBstWouldAddEndPuncttrue
	\mciteSetBstMidEndSepPunct{\mcitedefaultmidpunct}
	{\mcitedefaultendpunct}{\mcitedefaultseppunct}\relax
	\EndOfBibitem
	\bibitem[Str{\"u}mpfer and Schulten(2011)Str{\"u}mpfer, and
	Schulten]{Struempfer2011}
	Str{\"u}mpfer,~J.; Schulten,~K. The Effect of Correlated Bath Fluctuations on
	Exciton Transfer. \emph{The Journal of chemical physics} \textbf{2011},
	\emph{134}, 095102\relax
	\mciteBstWouldAddEndPuncttrue
	\mciteSetBstMidEndSepPunct{\mcitedefaultmidpunct}
	{\mcitedefaultendpunct}{\mcitedefaultseppunct}\relax
	\EndOfBibitem
	\bibitem[Nesterov and Berman(2015)Nesterov, and Berman]{Nesterov2015}
	Nesterov,~A.~I.; Berman,~G.~P. Role of Protein Fluctuation Correlations in
	Electron Transfer in Photosynthetic Complexes. \emph{Physical Review E}
	\textbf{2015}, \emph{91}, 042702\relax
	\mciteBstWouldAddEndPuncttrue
	\mciteSetBstMidEndSepPunct{\mcitedefaultmidpunct}
	{\mcitedefaultendpunct}{\mcitedefaultseppunct}\relax
	\EndOfBibitem
	\bibitem[Nazir(2009)]{Nazir2009}
	Nazir,~A. Correlation-Dependent Coherent to Incoherent Transitions in Resonant
	Energy Transfer Dynamics. \emph{Physical Review Letters} \textbf{2009},
	\emph{103}, 146404\relax
	\mciteBstWouldAddEndPuncttrue
	\mciteSetBstMidEndSepPunct{\mcitedefaultmidpunct}
	{\mcitedefaultendpunct}{\mcitedefaultseppunct}\relax
	\EndOfBibitem
	\bibitem[Merkli \latin{et~al.}(2016)Merkli, Berman, Sayre, Gnanakaran,
	K{\"o}nenberg, Nesterov, and Song]{Merkli2016}
	Merkli,~M.; Berman,~G.~P.; Sayre,~R.~T.; Gnanakaran,~S.; K{\"o}nenberg,~M.;
	Nesterov,~A.; Song,~H. Dynamics of a Chlorophyll Dimer in Collective and
	Local Thermal Environments. \emph{Journal of Mathematical Chemistry}
	\textbf{2016}, \emph{54}, 866--917\relax
	\mciteBstWouldAddEndPuncttrue
	\mciteSetBstMidEndSepPunct{\mcitedefaultmidpunct}
	{\mcitedefaultendpunct}{\mcitedefaultseppunct}\relax
	\EndOfBibitem
	\bibitem[Park and Galperin(2011)Park, and Galperin]{Park2011}
	Park,~T.-H.; Galperin,~M. Self-Consistent Full Counting Statistics of Inelastic
	Transport. \emph{Phys. Rev. B} \textbf{2011}, \emph{84}, 205450\relax
	\mciteBstWouldAddEndPuncttrue
	\mciteSetBstMidEndSepPunct{\mcitedefaultmidpunct}
	{\mcitedefaultendpunct}{\mcitedefaultseppunct}\relax
	\EndOfBibitem
	\bibitem[Ren \latin{et~al.}(2012)Ren, Zhu, Gubernatis, Wang, and Li]{Ren2012}
	Ren,~J.; Zhu,~J.-X.; Gubernatis,~J.~E.; Wang,~C.; Li,~B. Thermoelectric
	Transport with Electron-Phonon Coupling and Electron-Electron Interaction in
	Molecular Junctions. \emph{Phys. Rev. B} \textbf{2012}, \emph{85},
	155443\relax
	\mciteBstWouldAddEndPuncttrue
	\mciteSetBstMidEndSepPunct{\mcitedefaultmidpunct}
	{\mcitedefaultendpunct}{\mcitedefaultseppunct}\relax
	\EndOfBibitem
	\bibitem[Wang \latin{et~al.}(2012)Wang, Ren, Li, and Chen]{Wang2012a}
	Wang,~C.; Ren,~J.; Li,~B.; Chen,~Q.-H. Quantum Transport of Double Quantum Dots
	Coupled to an Oscillator in Arbitrary Strong Coupling Regime. \emph{The
		European Physical Journal B} \textbf{2012}, \emph{85}, 110\relax
	\mciteBstWouldAddEndPuncttrue
	\mciteSetBstMidEndSepPunct{\mcitedefaultmidpunct}
	{\mcitedefaultendpunct}{\mcitedefaultseppunct}\relax
	\EndOfBibitem
	\bibitem[Chen \latin{et~al.}(2008)Chen, Zhang, Liu, and Wang]{Chen2008}
	Chen,~Q.-H.; Zhang,~Y.-Y.; Liu,~T.; Wang,~K.-L. Numerically Exact Solution to
	the Finite-Size Dicke Model. \emph{Physical Review A} \textbf{2008},
	\emph{78}, 051801\relax
	\mciteBstWouldAddEndPuncttrue
	\mciteSetBstMidEndSepPunct{\mcitedefaultmidpunct}
	{\mcitedefaultendpunct}{\mcitedefaultseppunct}\relax
	\EndOfBibitem
	\bibitem[Mahan(2013)]{mahan2013many}
	Mahan,~G.~D. \emph{Many-Particle Physics}; Springer Science \& Business Media,
	2013\relax
	\mciteBstWouldAddEndPuncttrue
	\mciteSetBstMidEndSepPunct{\mcitedefaultmidpunct}
	{\mcitedefaultendpunct}{\mcitedefaultseppunct}\relax
	\EndOfBibitem
	\bibitem[Di~Ventra(2008)]{DiVentra2008}
	Di~Ventra,~M. \emph{Electrical Transport in Nanoscale Systems}; Cambridge
	University Press, 2008\relax
	\mciteBstWouldAddEndPuncttrue
	\mciteSetBstMidEndSepPunct{\mcitedefaultmidpunct}
	{\mcitedefaultendpunct}{\mcitedefaultseppunct}\relax
	\EndOfBibitem
	\bibitem[Meir and Wingreen(1992)Meir, and Wingreen]{Meir1992}
	Meir,~Y.; Wingreen,~N.~S. Landauer Formula for the Current Through an
	Interacting Electron Region. \emph{Phys. Rev. Lett.} \textbf{1992},
	\emph{68}, 2512--2515\relax
	\mciteBstWouldAddEndPuncttrue
	\mciteSetBstMidEndSepPunct{\mcitedefaultmidpunct}
	{\mcitedefaultendpunct}{\mcitedefaultseppunct}\relax
	\EndOfBibitem
	\bibitem[Pelzer \latin{et~al.}(2014)Pelzer, Can, Gray, Morr, and
	Engel]{Pelzer2014}
	Pelzer,~K.~M.; Can,~T.; Gray,~S.~K.; Morr,~D.~K.; Engel,~G.~S. Coherent
	Transport and Energy Flow Patterns in Photosynthesis under Incoherent
	Excitation. \emph{The Journal of Physical Chemistry B} \textbf{2014},
	\emph{118}, 2693--2702\relax
	\mciteBstWouldAddEndPuncttrue
	\mciteSetBstMidEndSepPunct{\mcitedefaultmidpunct}
	{\mcitedefaultendpunct}{\mcitedefaultseppunct}\relax
	\EndOfBibitem
	\bibitem[Cho \latin{et~al.}(2005)Cho, Vaswani, Brixner, Stenger, and
	Fleming]{Cho2005}
	Cho,~M.; Vaswani,~H.~M.; Brixner,~T.; Stenger,~J.; Fleming,~G.~R. Exciton
	Analysis in 2D Electronic Spectroscopy. \emph{The Journal of Physical
		Chemistry B} \textbf{2005}, \emph{109}, 10542--10556\relax
	\mciteBstWouldAddEndPuncttrue
	\mciteSetBstMidEndSepPunct{\mcitedefaultmidpunct}
	{\mcitedefaultendpunct}{\mcitedefaultseppunct}\relax
	\EndOfBibitem
	\bibitem[Mal{\`y} \latin{et~al.}(2016)Mal{\`y}, Somsen, Novoderezhkin,
	Man{\v{c}}al, and van Grondelle]{maly2016}
	Mal{\`y},~P.; Somsen,~O.~J.; Novoderezhkin,~V.~I.; Man{\v{c}}al,~T.; van
	Grondelle,~R. The Role of Resonant Vibrations in Electronic Energy Transfer.
	\emph{ChemPhysChem} \textbf{2016}, \emph{17}, 1356--1368\relax
	\mciteBstWouldAddEndPuncttrue
	\mciteSetBstMidEndSepPunct{\mcitedefaultmidpunct}
	{\mcitedefaultendpunct}{\mcitedefaultseppunct}\relax
	\EndOfBibitem
	\bibitem[Aghtar \latin{et~al.}(2014)Aghtar, Strümpfer, Olbrich, Schulten, and
	Kleinekathöfer]{Aghtar2014}
	Aghtar,~M.; Strümpfer,~J.; Olbrich,~C.; Schulten,~K.; Kleinekathöfer,~U.
	Different Types of Vibrations Interacting with Electronic Excitations in
	Phycoerythrin 545 and Fenna--Matthews--Olson Antenna Systems. \emph{The
		Journal of Physical Chemistry Letters} \textbf{2014}, \emph{5},
	3131--3137\relax
	\mciteBstWouldAddEndPuncttrue
	\mciteSetBstMidEndSepPunct{\mcitedefaultmidpunct}
	{\mcitedefaultendpunct}{\mcitedefaultseppunct}\relax
	\EndOfBibitem
	\bibitem[Tiwari \latin{et~al.}(2013)Tiwari, Peters, and Jonas]{tiwari2013}
	Tiwari,~V.; Peters,~W.~K.; Jonas,~D.~M. Electronic Resonance with
	Anticorrelated Pigment Vibrations Drives Photosynthetic Energy Transfer
	Outside the Adiabatic Framework. \emph{Proceedings of the National Academy of
		Sciences} \textbf{2013}, \emph{110}, 1203--1208\relax
	\mciteBstWouldAddEndPuncttrue
	\mciteSetBstMidEndSepPunct{\mcitedefaultmidpunct}
	{\mcitedefaultendpunct}{\mcitedefaultseppunct}\relax
	\EndOfBibitem
	\bibitem[Dijkstra \latin{et~al.}(2015)Dijkstra, Wang, Cao, and
	Fleming]{Dijkstra2015}
	Dijkstra,~A.~G.; Wang,~C.; Cao,~J.; Fleming,~G.~R. Coherent Exciton Dynamics in
	the Presence of Underdamped Vibrations. \emph{The journal of physical
		chemistry letters} \textbf{2015}, \emph{6}, 627--632\relax
	\mciteBstWouldAddEndPuncttrue
	\mciteSetBstMidEndSepPunct{\mcitedefaultmidpunct}
	{\mcitedefaultendpunct}{\mcitedefaultseppunct}\relax
	\EndOfBibitem
	\bibitem[Novoderezhkin \latin{et~al.}(2017)Novoderezhkin, Romero, Prior, and
	van Grondelle]{Novoderezhkin2017}
	Novoderezhkin,~V.~I.; Romero,~E.; Prior,~J.; van Grondelle,~R.
	Exciton-Vibrational Resonance and Dynamics of Charge Separation in the
	Photosystem II Reaction Center. \emph{Physical Chemistry Chemical Physics}
	\textbf{2017}, \emph{19}, 5195--5208\relax
	\mciteBstWouldAddEndPuncttrue
	\mciteSetBstMidEndSepPunct{\mcitedefaultmidpunct}
	{\mcitedefaultendpunct}{\mcitedefaultseppunct}\relax
	\EndOfBibitem
\end{mcitethebibliography}

\providecommand{\latin}[1]{#1}
\makeatletter
\providecommand{\doi}
{\begingroup\let\do\@makeother\dospecials
	\catcode`\{=1 \catcode`\}=2 \doi@aux}
\providecommand{\doi@aux}[1]{\endgroup\texttt{#1}}
\makeatother
\providecommand*\mcitethebibliography{\thebibliography}
\csname @ifundefined\endcsname{endmcitethebibliography}
{\let\endmcitethebibliography\endthebibliography}{}

\end{document}